# Anisotropic spin splitting of the electron ground state in InAs quantum dots

E. Aubry, C. Testelin, F. Bernardot, and M. Chamarro
*Institut des NanoSciences de Paris, Universités Paris-VI et Paris-VII,*
*CNRS UMR 7588, Campus Boucicaut, 140 rue de Lourmel, 75015 Paris, France*

A. Lemaître
*CNRS – Laboratoire de Photonique et Nanostructures, Route de Nozay,*
*91460 Marcoussis, France*

ABSTRACT

Photoinduced circular dichroism experiments in an oblique magnetic field allow measurements of Larmor precession frequencies, and so give a precise determination of the electron Landé $g$ factor and its anisotropy in self-assembled InAs/GaAs quantum dots emitting at 1.32 eV. In good agreement with recent theoretical results, we measure $|g_\perp| = 0.397 \pm 0.003$ and $|g_{//}| = 0.18 \pm 0.02$.

The spin of an individual carrier spatially localized in a quantum dot (QD) is a promising candidate for implementation of qu-bits in view of applications in the fields of spintronics and quantum information. In this direction, an expected and very attractive property of electronic spin is its long stability in time. Indeed, the localization and confinement should lead to the suppression of spin-flip mechanisms active in bulk and 2D systems. In last years pump-probe photoluminescence measurements in an ensemble of QDs gave a single-electron spin relaxation time of 15 ns[1], and photoinduced Faraday rotation experiments gave a single-electron spin coherence time of 3 µs[2] (at B = 6 T, T = 10 K). Longer spin relaxation times, in the ms range, have been also found for an electronic spin placed in a longitudinal magnetic field (4 and 8 T), at low temperatures[3,4]. Nowadays, large experimental spin decoherence times make possible to perform optical manipulations while spin coherence is retained. Optical manipulation of such spins is an area of increasing interest and, recently, several theoretical schemes have been formulated in presence of a static magnetic field[5,6]. The optical manipulations are then governed by the value of the Landé $g$ factor and its anisotropic properties. In addition, the anisotropic part of $g$ influences spin-lattice relaxation[7] and plays an important role in spin-related applications.

In this paper, we report on the precise determination of the electron Landé $g$ factor and its anisotropy in self-assembled InAs/GaAs QDs, by means of measurements of Larmor precession frequencies in photoinduced circular dichroism (PCD) experiments performed with an oblique magnetic field. In the past, two main kinds of experiments have been performed to obtain the electronic $g$ factor in QDs: magneto-microluminescence[8,9,10,11,12,13,14] or magneto-capacitance[15,16,17] measurements. Magneto-microluminescence gives an effective $g$ factor that contains both electron and hole contributions; the analysis of the dependence of the electron-hole pair energy on magnetic field leads to separation of both contributions, and implies the detection of isolated QDs to avoid inhomogeneous broadening. Here, we use an optical technique that, applied to an ensemble of QDs, is able to measure the electronic $g$ factor corresponding to ground-state electrons resident in QDs emitting at a given energy of fundamental optical transition.

We have chosen to focus our work on small self-assembled InAs/GaAs QDs because, despite the fact that important experimental results have been obtained in last years in these materials[18,19,20,21], there are only few experimental informations about their magnetic behaviour. However, an important theoretical effort has been recently developed to understand the size dependence[22,23] of carriers Landé g factor and its anisotropy[24,25]. In particular, a very detailed theoretical work[23] has predicted an important size dependence of the electron g factor parallel to the growth axis, $g_{//}$, and a quasi-constant value of the electron g factor perpendicular to the growth axis, $g_{\perp}$, leading to an anisotropy ratio of the Landé g factor (defined as $g_{//}/g_{\perp}$) varying from roughly three to almost zero when the emission energy goes from 1.05 eV to 1.30 eV. Previous measurements of g anisotropy were performed in $In_{0.6}Ga_{0.4}As$ self-assembled QDs[8] or more recently in large InAs QDs[17]. Meanwhile, other groups have found isotropic electronic g factors in self-assembled InAs[13] or GaAs[12] QDs. We show that our experimental determination of the electron g factors is in very good agreement with theoretical predictions, and adds to the size dependence of the avaible experimental data.

The studied structure was grown by molecular beam epitaxy on a (001) GaAs substrate. It consists of 30 planes of InAs self-assembled QDs, separated by 40-nm thick GaAs spacer layers. The plane QD density is about $10^{10}$ cm$^{-2}$. For an average occupation by a single electron per dot, the structure was n-modulation doped 2 nm below each layer with a Si-dopant density equal to 6 10$^{10}$ cm$^{-2}$. The photoluminescence spectrum at 2 K shows a maximum at 1.31 eV and a full width at half maximum of about 80 meV.

To perform PCD measurements, a picosecond Ti:sapphire laser tuned to 1.32 eV is split into pump and probe beams. The pump beam polarization is σ+/σ− modulated at 42 KHz with a photoelastic modulator; the probe beam is linearly polarized. After transmission through the sample, the probe beam is spatially resolved into its two circular components, and the difference in their intensities is measured in a balanced optical bridge. To improve signal-to-noise ratio, a double lockin amplifier is used for the measurement, pump and probe beams being modulated with an optical chopper at two different frequencies.

Figure 1 (a) shows the PCD signal obtained at 2 K and zero magnetic field, *versus* pump-probe delay time. A fast decay in the first hundreds of picoseconds is observed, followed by a slower decay. The initial fast decay is due to the presence in the sample of undoped QDs, for which the experimental configuration allows the measurement of a dynamical dichroism signal[21]. The lowest optically excited state of n-doped QDs is a trion state, consisting of a singlet pair of electrons and a heavy hole with its spin pointing up or down depending on the σ+ or σ− circularly polarized light of excitation. Assuming a very slow spin relaxation time for holes[26], the PCD signal associated to singly charged QDs is written as a two-exponential decay:

$$PCD\left[\vec{B}=\vec{0}\right] \propto \frac{2T_1 - T_R}{T_1 - T_R} e^{-t/T_R} - \frac{T_R}{T_1 - T_R} e^{-t/T_1} , \qquad (1)$$

where $T_R$ is the trion recombination time and $T_1$ is the spin relaxation time of electrons. If $T_1 \gg T_R$, as has been experimentally found recently[27], the PCD signal is described by a single exponential with characteristic time $T_R$. Fit of our PCD decay curve gives a time of about 600 ps, which is in good agreement with trion recombination times obtained in n-doped InAs QDs[27].

The sample was placed in a cryostat containing superconducting coils. Figures 1 (b), (c) and (d) show PCD signals obtained when the applied magnetic field $\vec{B}$ is perpendicular to both pump and probe directions, *i.e.* parallel to the QDs planes. The initial fast decay related to empty QDs is always observed in these curves. We also remark an oscillatory signal, which is not symmetric with respect to the zero signal level. We use a moderate value of B (B < 1.5 T) in order to avoid the observation of a supplementary oscillatory signal associated to the precession of the photo-excited hole during the lifetime of trions[8]. Under these conditions, in an applied magnetic field, the PCD signal of n-doped QDs can be described by two terms[28]: a non-oscillatory term described by an exponential decay, associated to the polarized holes contained in trion complexes, and a second term associated to the damped precession of the oriented electron spins around the magnetic field, with Larmor frequency $\omega_e$:

$$PCD\left[\vec{B} \neq \vec{0}\right] \propto A e^{-t/T_R} + B e^{-t/T_2^*} \cos(\omega_e t - \delta) , \qquad (2)$$

where $T_2^*$ is the decoherence time of the electronic spins. By fitting our experimental curves to Eq. (2), we have obtained $T_R$, $\omega_e$ and $T_2^*$. We found $T_R = 700 \pm 100$ ps, which are values very similar to the value obtained at B = 0. The first term of Eq. (2) is at the origin of the signal asymmetry. As expected, the oscillation frequency increases linearly with magnetic field: $\omega_e = g_\perp \mu_B B / \hbar$; inset in Fig. 2 shows a linear fit from which we have determined $|g_\perp| = 0.397 \pm 0.003$. $T_2^*$ is found about 1 ns for 0.75 T < B < 1.25 T. Recent experiments on ensembles of QDs show a dependence of $T_2^*$ on magnetic field[27,28], $T_2^*$ decreasing for increasing magnetic fields. That was explained by an inhomogenity of *g* values among the observed QDs. Our experimental window for pump-probe delay and the moderate values of applied magnetic field, lead to a lack of precession for the observation of a dependence of $T_2^*$ on magnetic field.

The sample can also be rotated about an axis perpendicular to $\vec{B}$ and the laser beams. The rotation angle α is measured between $\vec{B}$ and the sample plane direction. β is the angle which appears when α ≠ 0, and is defined as the angle between the growth direction and the direction of the oriented electronic spin given by $\vec{S}$ in the inset of Fig. 3. For an anisotropic g tensor, the spin-precession axis $\vec{\Omega}$ is, generally, non-colinear with $\vec{B}$. Axis $\vec{\Omega}$ is tilted away from the direction of $\vec{B}$ by an angle γ, given by $\tan\gamma = (g_{//}/g_\perp)\tan\alpha$. The electronic spin $\vec{S}$ contains precessing and non-precessing components, which correspond, respectively, to the perpendicular and parallel projections of $\vec{S}$ onto $\vec{\Omega}$ axis.

Figure 3 shows PCD measurements at B = 0.86 T, for α = 0° and α = 30°. At α ≠ 0, the oscillating signal is superimposed on a nonoscillating signal which has two different origins: the first one, associated to the trions, was already described for α = 0 and decreases with characteristic time $T_R$; the second one is due to the nonprecessing electron spin component, along the Ω axis (see inset of Fig. 3), and is damped with characteristic time $T_1$. As demonstrated in Fig. 3, the frequency of oscillation disminishes when α increases. The effective g factor for a tilted magnetic field, when α ≠ 0, is given by:

$$g_{eff} = \sqrt{g_\perp^2 \cos^2\alpha + g_{//}^2 \sin^2\alpha} \ . \tag{3}$$

We have measured $g_{eff}$, at each explored value of α, by a linear fit of the oscillation frequency *versus* magnetic field, as we made for $|g_\perp|$. The $g_{eff}$ values measured in this procedure are represented in Fig. 2, *versus* α angle. We finally obtain, through a fit to Eq. (3), $|g_{//}| = 0.18 \pm 0.02$.

We propose to use Pryor and Flatté's recent theoretical results[23] as a general framework to understand the ensemble of experimental values of electronic g factors for different QD sizes. Indeed, the $|g_\perp|$ and $|g_{//}|$ values that we have obtained in InAs/GaAs QDs for an energy of the lowest optical transition of 1.32 eV are in very good agreement with this work, and adds to experimental data obtained in larger InAs QDs[17] emitting at 1.05 eV, which are also in good agreement with the same calculations ($|g_\perp| = 0.57$ and $|g_{//}| = 1.51$). We notice that, as predicted, $|g_\perp| > |g_{//}|$ for smaller sizes and $|g_\perp| < |g_{//}|$ for larger sizes. At intermediate QD sizes, the electronic g factor should be isotropic; this phenomenon has been observed in neutral InGaAs QDs, on which isotropic[13] ($|g_\perp| = |g_{//}| = 0.55$) or quasi-isotropic[14] ($|g_\perp| = 0.65$, $|g_{//}| = 0.80$) electron g factors have been measured[30,29].

Concerning the hole g factor, the absence of supplementary modulation on our PCD signals, during the trion lifetime and for all α values, seems to indicate a very small hole $g_\perp^h$ factor, in agreement with the calculations presented in Ref. 23. Nonetheless, taking into account the present and previous[30] experimental results on n-doped InAs QDs emitting in the 1.3 eV region, we can estimate the hole $g_{//}^h$ factor to be $|g_{//}^h| \approx 2.5$. This result indicates that $|g_{//}^h|$ is theoretically overestimated in Ref. 23 ($g_{//}^h \approx 8$), and that more experimental and theoretical studies of the hole *g* factor in InAs QDs are still needed.

**Figure captions**

**Figure 1** PCD signals of n-doped InAs/GaAs QDs, obtained at 2 K. Pump and probe were tuned to the first optical transition energy, at 1.32 eV. Upper curve was obtained with no applied magnetic field, and other curves were obtained for several values of an applied magnetic field perpendicular to the growth axis of the sample (Voigt configuration). Dashed horizontal lines represent zero signals for each curve.

**Figure 2** Inset: Linear field dependence of the electron Larmor frequency at $\alpha = 0$ (see inset of Fig. 3), from which we get $|g_\perp| = 0.397 \pm 0.003$. Effective Landé factor *versus* tilt angle $\alpha$; the full squares represent experimental data, and the solid line is a fit to Eq. (3) from which we obtain $|g_{//}| = 0.18 \pm 0.02$.

**Figure 3** PCD signals, measured at 2 K, of n-doped InAs/GaAs QDs in presence of an applied magnetic field of 0.85 T, for two different angles $\alpha$ (0° and 30°). Pump and probe were tuned to the first optical transition energy, at 1.32 eV. Inset represents the geometry of application of a tilted magnetic field: the plane of the sample makes angle $\alpha$ with the magnetic field $\vec{B}$; the initial electronic spin $\vec{S}$ makes angle $\beta$ with the growth axis z; the spin rotation axis $\vec{\Omega}$ and the magnetic field are misaligned by angle $\gamma$.

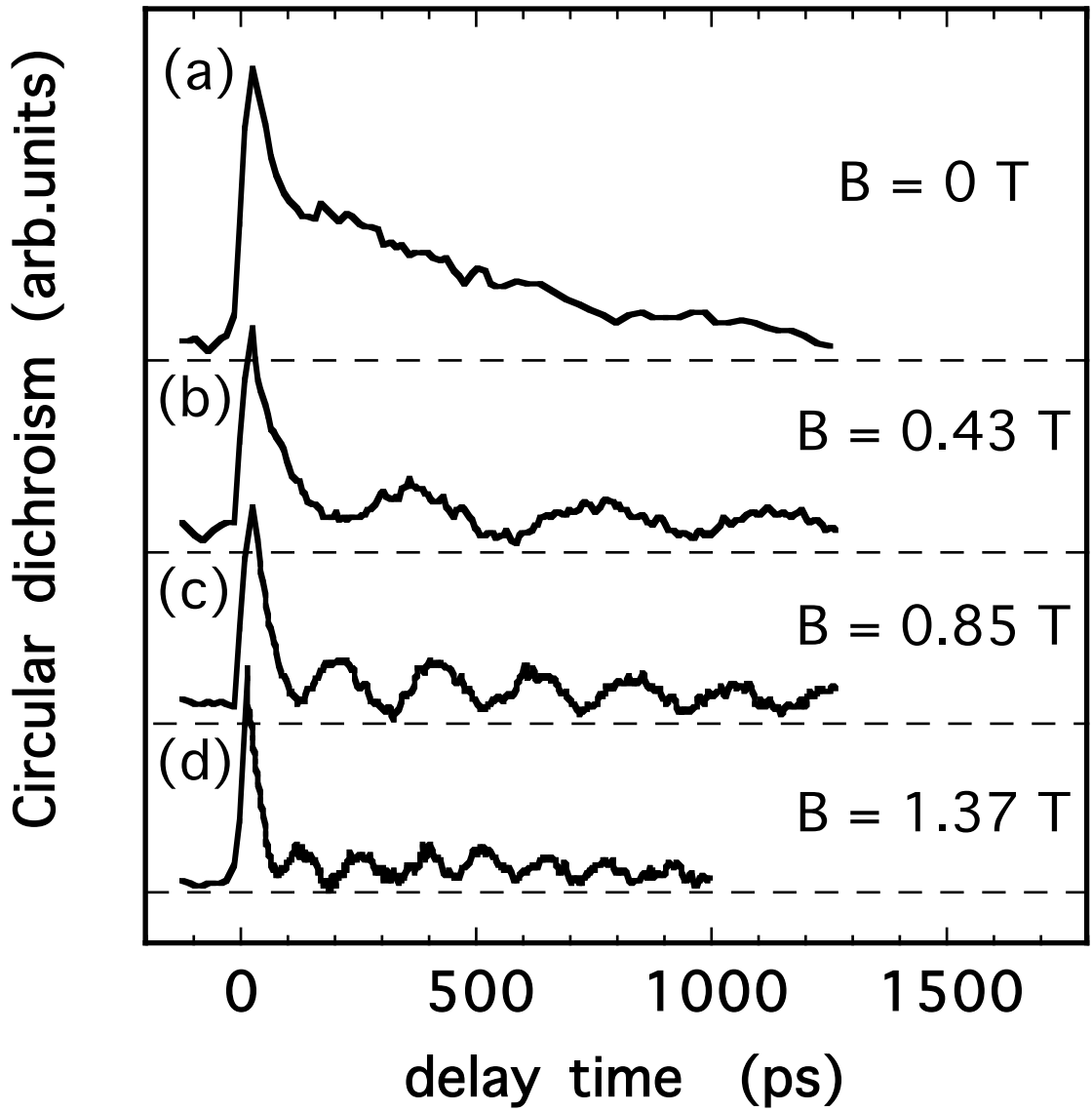

Figure 1

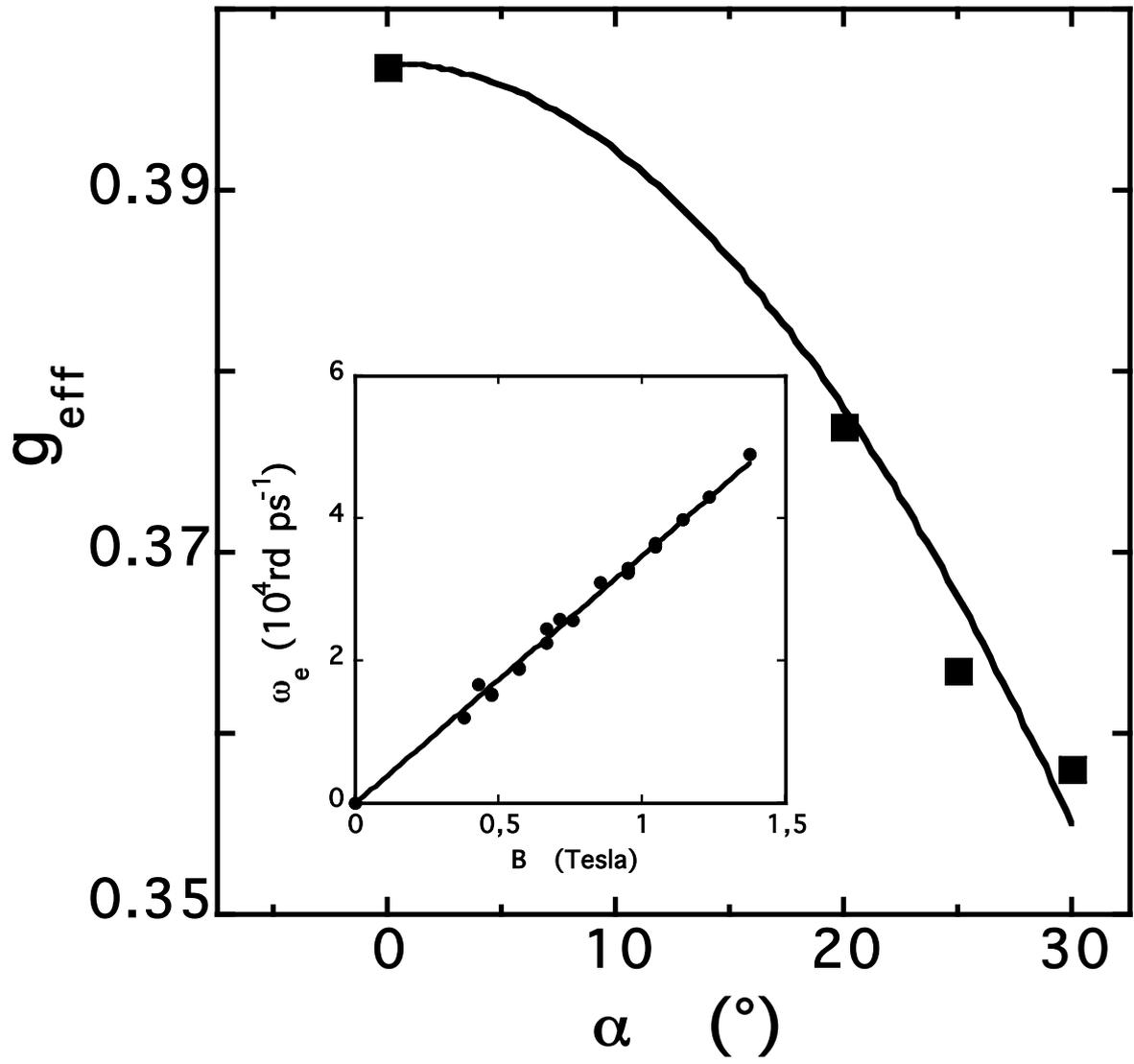

Figure 2

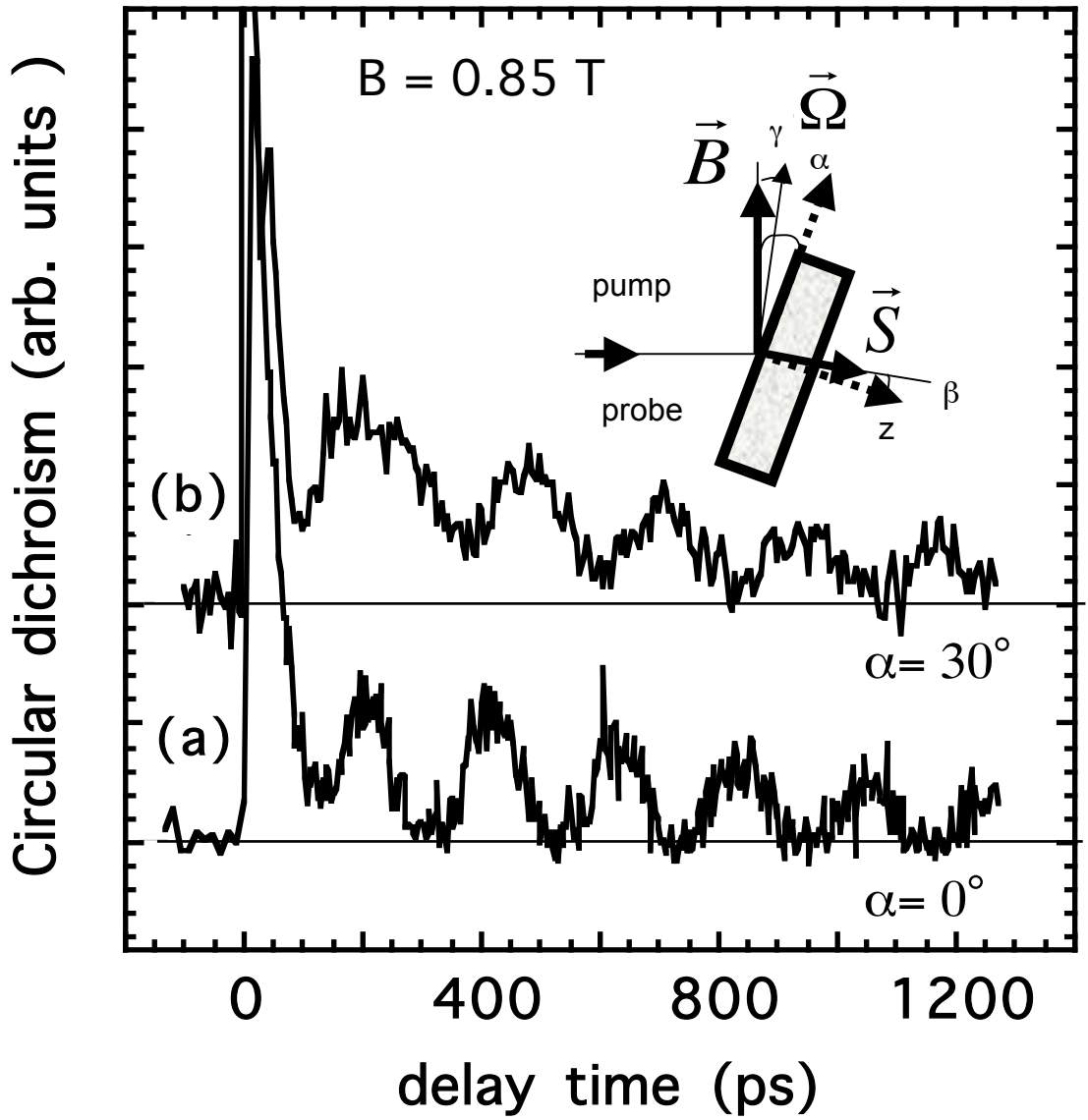

Figure 3